\documentclass[preprint,5p,times,twocolumn,10pt]{elsarticle}

\journal{Atomic Data and Nuclear Data Tables}

\usepackage[utf8]{inputenc}
\usepackage{amsmath, amssymb}
\usepackage{graphicx}
\usepackage{siunitx}
\usepackage{booktabs}
\usepackage{hyperref}
\usepackage{geometry}
\newcommand{\Wtrois}{$^\text{183}$W}
\newcommand{\Wquatre}{$^\text{184}$W}
\newcommand{\nxn}{(n,~$x$n)}
\newcommand{\nxng}{(n,~$x$n~$\gamma$)}
\newcommand{\ndeuxn}{(n,~2n)}
\newcommand{\ndeuxng}{(n,~2n~$\gamma$)}
\newcommand{\ninel}{(n,~n$'$)}
\newcommand{\ninelg}{(n,~n$'$~$\gamma$)}
\newcommand{\ntroisn}{(n,~3n)}

\begin{document}

\sloppy

\begin{frontmatter}

\title{Experimental \Wquatre\ndeuxng{}~reaction cross sections} 

\author[iphc]{Greg~Henning\corref{cor1}}
\ead{ghenning@iphc.cnrs.fr}
\cortext[cor1]{Corresponding author}
\affiliation[iphc]{organization={Universit\'e de Strasbourg, Centre National de la Recherche Scientifique, IPHC UMR 7178},
            addressline={23 rue du Loess}, 
            city={Strasbourg},
            postcode={BP28 67037}, 
            country={France}}
            
\author[iphc]{Antoine~Bacquias}
\author[des]{David~Bernard}
\author[ifin]{Catalin~Borcea}
\author[ifin]{Marian~Boromiza}
\author[aiea]{Roberto~Capote}
\author[ifin]{Adina~Coman}
\author[dam,UOrsay]{Cyrille~de~Saint-Jean}
\author[iphc]{Philippe~Dessagne}
\author[dam,UOrsay]{Marc~Dupuis}
\author[jrc]{Jean-Claude~Droh\'e}
\author[oakridge]{Klaus~H.~Guber}
\author[iphc]{Ma\"elle~Kerveno}
\author[iphc]{G\'erard~Rudolf\fnref{retired}}
\fntext[retired]{Retired}
\author[iphc]{Pol~Scholtes}
\author[dam,UOrsay]{St\'ephane~Hilaire}
\author[dam,UOrsay]{Pascal~Romain}
\author[losalamos]{Toshihiko~Kawano}
\author[ifin]{Andu~Negret}
\author[des]{Gilles~Noguere}
\author[jrc]{Markus~Nyman\fnref{helsinki}}
\fntext[helsinki]{Currently, Univ. of Helsinki}
\author[jrc]{Andreea~Oprea}
\author[jrc]{Carlos~Paradela}
\author[jrc]{Arjan~J.M.~Plompen}
\author[iphc]{Andrea~Tsinganis}

\affiliation[des]{organization={CEA, DES, IRESNE},
            city={Saint Paul l\`es Durance},
            postcode={13108}, 
            country={France}}

\affiliation[ifin]{organization={Horia Hulubei National Institute for Physics and Nuclear Engineering},
            city={Magurele},
            country={Romania}}

\affiliation[aiea]{organization={Nuclear Data Section, International Atomic Energy Agency},
            addressline={Wagramer Strasse}, 
            city={Vienna},
            postcode={1400}, 
            country={Austria}}

\affiliation[dam]{organization={CEA, DAM, DIF},
            city={Arpajon},
            postcode={912397}, 
            country={France}}
\affiliation[UOrsay]{organization={Universit\'e Paris-Saclay, CEA, Laboratoire Mati\`ere sous Conditions Extr\^emes},
            city={Bruy\`eres-le-Ch\^atel},
            postcode={91680},
            country={France}}
\affiliation[jrc]{organization={European Commission, Joint Research Centre},
            city={Geel},
            postcode={2440}, 
            country={Belgium}}

\affiliation[losalamos]{organization={Los Alamos National Laboratory},
            city={Los Alamos},
            postcode={87545}, 
            state={New Mexico},
            country={USA}}
            
\affiliation[oakridge]{organization={Oak Ridge National Laboratory},
            city={Oak Ridge},
            postcode={37831}, 
            state={TN},
            country={USA}}

\begin{abstract}
This paper presents new experimental cross sections for the $^{184}$W(n, 2n~$\gamma$)$^{183}$W reaction, providing essential data to improve evaluated nuclear databases used in the design of advanced fission and fusion reactors. 
The measurements were performed at the GELINA neutron beam facility (JRC-Geel, Belgium) using the GRAPhEME setup, which consists of four high-purity planar Ge detectors and a $^{235}$U fission chamber for incident neutron flux monitoring. 
Using an isotopically enriched $^{184}$W target, the experimental method combines prompt $\gamma$-ray spectroscopy with the neutron time-of-flight technique to determine incident neutron energies, to extract reaction cross section from threshold ($\approx$ \SI{7.5}{MeV}) up to approximately \SI{25}{MeV}. 
Partial $\gamma$-ray production cross sections were determined for six discrete transitions. 
The data analysis relies on a Monte Carlo approach for the propagation of systematic and statistical uncertainties. 
Results are compared with theoretical predictions from the TALYS reaction code.
\end{abstract}



\end{frontmatter}

\section{Introduction}

The development and simulation of advanced nuclear reactors rely fundamentally on high-precision nuclear data. 
Evaluated databases~\cite{ENDFB8,JEFF33,JENDL5} are a fundamental input for these numerical simulations, yet the current level of uncertainty in these evaluations still limits the ability of modern calculations to reach the required level of predictive accuracy~\cite{SWG26,SG46Cabellos2023,SG46Cabellos2024}.
To improve the evaluations of nuclear data, it is essential to acquire new, high-quality experimental measurements, and to refine the underlying theoretical descriptions of the nuclear reactions involved.

Tungsten ($_{74}$W) isotopes are of particular relevance for a range of nuclear applications. 
They are integral components in alloys designed for both fission and fusion reactors, prized for their high melting point, strong mechanical resistance, and low thermal expansion~\cite{CRCHAndbook2016}. 
From a nuclear physics perspective, tungsten isotopes offer a unique opportunity for research. 
Their nuclear structure is characterized by a strong deformation, similar to that of actinides~\cite{MOLLER1995185}, but they lack a neutron-induced fission channel in the energy range of interest. 
This makes them ideal systems for benchmarking and refining nuclear reaction models without the added complexity of fission.

Despite their importance, a significant gap exists in the experimental data for tungsten. Prior to this work, public databases like the Experimental Nuclear Reaction Data library (EXFOR ~\cite{EXFOR}) contained no measurements for the total \Wquatre\ninel{} cross-section (with existing data limited to a few specific level-production cross-sections) and only a sparse dataset for the \ndeuxn{} channel, often related to isomer production.
This scarcity of data limits the ability to validate and constrain theoretical models.
In particular, $\gamma$ rays emitted following \ndeuxn{} reactions are a key probe of the pre-equilibrium regime in neutron-induced reactions, and are sensitive to the structure of the residual nucleus.

The primary objective of this paper is to address this gap by presenting new experimental data for the \Wquatre\ndeuxng{} reaction.
We report the cross sections for a set of six discrete $\gamma$-ray transitions produced in this reaction channel.
In Ref.~\cite{EPJA183W}, the cross sections of the same six low-lying transitions of \Wtrois{} (below \SI{500}{keV}) were measured in the \Wtrois\ninelg{} channel with full covariance matrices, and the comparison with TALYS calculations showed that the magnitude and global shape of the cross sections are reasonably described, but with significant discrepancies depending on the transition and on the model parametrization.
The present measurement therefore provides the same \Wtrois{} transitions populated through the \Wquatre\ndeuxn{} reaction, so that the two sets can be used together to test the models on the same final nucleus.
The \nxng{} measurement technique, which combines prompt $\gamma$-ray spectroscopy with the neutron time-of-flight method, is an indirect approach to \nxn{} cross sections; the conditions required to obtain accurate cross sections with it are discussed in Ref.~\cite{Kerveno2018EPJN}.
Partial $\gamma$-ray production cross sections are a stringent test of reaction models, as they are sensitive to both the reaction mechanism and the structure of the residual nucleus~\cite{Kerveno2020ND2019,Kerveno2025CNR24}.
The performance and main achievements of the GRAPhEME setup at GELINA are reviewed in Ref.~\cite{Grapheme2023}.
The following sections detail the experimental setup, the data analysis, and the results.

\section{Experimental Setup}



The experiment was performed at the GELINA (Geel Electron Linear Accelerator) facility, operated by the European Commission's Joint Research Centre (EC-JRC) in Geel, Belgium~\cite{GelinaWeb,Gelina1,Gelina2}. 
GELINA provides a pulsed white neutron beam (repetition rate of \SI{800}{Hz} at the time of the experiment), with a continuous energy spectrum extending from the keV range up to approximately \SI{25}{MeV}, which makes it well suited to studying neutron-induced reactions in the fast-neutron range, including the \ndeuxn{} channel.
The GRAPhEME (GeRmanium array for Actinides PrEcise MEasurements) setup was used to perform the measurements~\cite{thiry2010,7465505,Kerveno2015Grapheme,Grapheme2023}. 
This system is specifically designed for the high-resolution detection of $\gamma$-rays emitted following neutron-induced reactions. 
It is positioned at a $28.82 \pm 0.01$ meters distance from the neutron production target. This distance allows for the precise determination of the incident neutron energy using the time-of-flight (ToF) technique, where the time difference between the neutron pulse and the detection of a $\gamma$-ray is measured.
A $^{235}$U fission chamber (FC) is placed in the neutron beam just upstream of the tungsten target. This chamber continuously measures the incoming neutron flux, providing the absolute normalization required to convert the measured $\gamma$-ray yields into absolute reaction cross-sections.
An array of four planar High Purity Germanium (HPGe) detectors is arranged around the target. 
Planar detectors were chosen for their excellent energy resolution at the lower $\gamma$-ray energies characteristic for this reaction channel,
which is essential for cleanly resolving discrete transitions and minimizing peak overlap.
A fully digital acquisition system records the energy deposited in each detector and the precise timing of each event in list mode. 
This comprehensive data stream allows for detailed offline analysis of $\gamma$ energy vs. time-of-flight spectra.
%
Finally, a metallic tungsten target, isotopically enriched to \SI{94.5}{\percent} in \Wquatre{}, is placed at the center of the setup.
This high enrichment ensures that the measured $\gamma$ rays can be attributed to reactions on \Wquatre{}.
The other tungsten isotopes each account for less than \SI{1}{\percent}; the remainder consists of other impurities, including oxygen, whose exact composition is not known.
The target has a mass of \SI{45.280 \pm 0.015}{g} and a thickness of \SI{1.14 \pm 0.01}{mm}.
The target diameter is \SI{7.1}{cm}, while the neutron beam spot at the target position is $\approx$\SI{5}{cm}. 

\section{Data Analysis}
%
The analysis begins by identifying the characteristic $\gamma$-ray peaks in the energy spectra recorded by the HPGe detectors. 
The intensity of a specific $\gamma$-ray line is determined for a series of incident neutron energy windows, which are defined by binning the time-of-flight data. 
The neutron energy is determined as 
$$E_\text{n} = m_\text{n} c^2 \left( \left[1 - \left(\frac{DoF / ToF}{c}\right)^2\right]^{-1/2} - 1 \right)$$

(where $m_\text{n}$ is the mass of the neutron in $\text{MeV} / \text{c}^2$, $DoF$ and $ToF$ the distance and time of flight, respectively).

The number of fission events detected in the FC $N_\text{FC}(E_\text{n})$ is used to calculate the neutron flux after correcting for fragment detection efficiency ($\varepsilon_\text{FC} = 94.4\pm2.1$\% \cite{thiry2010}), for the amount of $^\text{235}$U in the deposit $N_{^{235}U}$ ($324 \pm 2 ~\mu\text{g}/\text{cm}^2$), and dividing by the fission induced cross section $\sigma_{^{235}\text{U(n,~f)}}(E_n)$ from ENDF/B-VIII.0~\cite{ENDFB8}. 
The number of neutrons going through the fission chamber is calculated to be:
$$N_\text{n} (E_\text{n}) = \frac{1}{N_{^{235}U}} \frac{N_\text{FC}(E_\text{n})}{\varepsilon_\text{FC}} \frac{1}{\sigma_{^{235}\text{U(n,f)}}(E_n)}$$

The number of detected $\gamma$ rays for specific neutron energies, $N_\gamma^\theta(E_\text{n})$, for a given HPGe detector at the angle $\theta$ combined with the number of neutrons going through the target $N_\text{n} (E_\text{n})$, determined with the fission chamber, and normalization factors (the $\gamma$ detection efficiency $\varepsilon(E_\gamma)$ and the amount of material in the target $N_\text{target}$) are used to obtain the partial cross-sections at the angle of the detector ${\frac{d\sigma_{\gamma}^{\theta}}{d\Omega}}(E_\text{n})$.

This process yields the angular differential cross-section $\mathrm{d}\sigma/\mathrm{d}\Omega$ for each detector angle: 
$$
{\frac{d\sigma_{\gamma}^{\theta}}{d\Omega}}(E_\text{n}) =
   \frac{1}{4 \pi} ~
   \frac{1}{N_\text{target}} ~
   \frac{{N_\gamma^\theta}({E_n})}{\varepsilon_{E_\gamma}} ~
   \frac{1}{{N_\text{n}}(E_\text{n})}
$$

To obtain the total, angle-integrated production cross-section $\sigma_{\gamma\text{total}}$ for each $\gamma$-ray, the results from detectors at two specific angles ($110^\circ$ and $150^\circ$), chosen to optimize the angle integration using the Gaussian quadrature method~\cite{Brune2002}, are combined  as follow 
$$\sigma_{\gamma}(E_\text{n}) = 4 \pi \left( w_{110^{\circ}} \frac{d\sigma_{\gamma}^{110}}{d\Omega}(E_\text{n}) +  w_{150^{\circ}} \frac{d\sigma_{\gamma}^{150}}{d\Omega}(E_\text{n}) \right) $$

with $w_{110^{\circ}} = 0.6521$ and $w_{150^{\circ}} = 0.3479$.

A full Monte Carlo analysis, similar to that used in Ref.~\cite{PRCeveneven}, was performed to quantify the final uncertainties.
For each iteration, all parameters involved in the calculation (e.g., detector efficiencies, target mass, timing resolution, neutron flux normalization) are randomly varied according to their probability distributions (Gaussian or uniform). The quoted uncertainties correspond to one standard deviation of the resulting distributions.
Potential sources of systematic bias or uncertainties were investigated and quantified, including timing degradation for low-energy $\gamma$-rays and multiple neutron scattering in the target. 
These effects were incorporated into the Monte Carlo uncertainty propagation, as described in Refs.~\cite{Henning2023MC,PRCeveneven}.

\section{Results}

We extracted the cross sections for six $\gamma$ rays emitted by \Wtrois{} nuclei produced in \Wquatre\ndeuxn{} reactions.
These $\gamma$ rays were selected because they correspond to the transitions studied in the \Wtrois\ninelg{} channel in Ref.~\cite{EPJA183W}.
Table~\ref{tab:transitions} summarizes the transitions for which cross-sections were measured.

\begin{table*}[!ht]
    \centering
    \begin{tabular}{l|ll|ll|c|r}
        ~ & \multicolumn{2}{c}{Initial Level} & \multicolumn{2}{c}{Final Level} &  &  \\ 
        $\gamma$ Energy  & Excitation & Spin  & Excitation  & Spin & Multipolarity & \multicolumn{1}{c}{ICC}  \\ 
        {[keV]} & Energy & Parity & Energy & Parity & &   \\ 
         & [keV] & ~ & [keV] & ~ & ~ & ~ \\ \hline
        \hline
        209.8 & 308.9 & 9/2$^{-}$ & 99.1 & 5/2$^{-}$ & E2 & 0.262 \\
        259.5 & 551.2 & 9/2$^{-}$ & 291.7 & 5/2$^{-}$ & [E2] & 0.133 \\
        268.1 & 475.1 & 11/2$^{-}$ & 207.0 & 7/2$^{-}$ & [E2] & 0.119 \\ 
        291.7 & 291.7 & 5/2$^{-}$ & 0 & 1/2$^{-}$ & E2 & 0.092 \\ 
        313.0 & 412.1 & 7/2$^{-}$ & 99.1 & 5/2$^{-}$ & M1+E2 {\footnotesize ($\delta=+0.225$)} & 0.190 \\ 
        353.9 & 453.1 & 7/2$^{-}$ & 99.1 & 5/2$^{-}$ & M1+E2 {\footnotesize ($\delta=-0.192$)} & 0.137   \\ \hline
    \end{tabular}
    
    \caption{Table of the $\gamma$-ray transitions in \Wtrois{} studied in this work for the \ndeuxng{} channel, ordered by $\gamma$-ray energy $E_\gamma$. 
    $E_\text{x}$ and $J^\pi$ are the excitation energy and spin-parity of the initial and final levels.
    Structure information was extracted from Refs.~\cite{nudat,BAGLIN2016149}; in particular, multipolarities are quoted as in the ENSDF evaluation distributed by the National Nuclear Data Center (NNDC), square brackets denoting an assumed multipolarity.
    The total internal conversion coefficients, $\alpha_\text{tot}$, are taken from the TALYS structure input files (which usually quote the RIPL-3 database~\cite{RIPL3}) and were checked with the BrIcc (Band-Raman internal conversion coefficients) calculator~\cite{KIBEDI2008202}.
    For transitions of mixed multipolarity, the mixing ratio $\delta$ is given when known.
    }
    \label{tab:transitions}
\end{table*}

The measured cross sections all display a threshold around \SI{7.5}{MeV} (consistent with the \SI{7.411}{MeV} neutron separation energy of \Wquatre{}) and a maximum around \SIrange{12}{15}{MeV} (corresponding to the opening of the \Wquatre\ntroisn{} channel at \SI{13.6}{MeV}).

Tables \ref{tab:209}, \ref{tab:259}, \ref{tab:268}, \ref{tab:291}, \ref{tab:313}, and \ref{tab:354} give all the values extracted from the analysis.
The rounding and truncation of numbers may lead to apparent discrepancies between uncertainties columns. Refer to the data repository (Ref.~\cite{QFFRD8_2025}) for the full numbers).

\begin{figure*}[!ht]
\centering
\includegraphics[scale=.41]{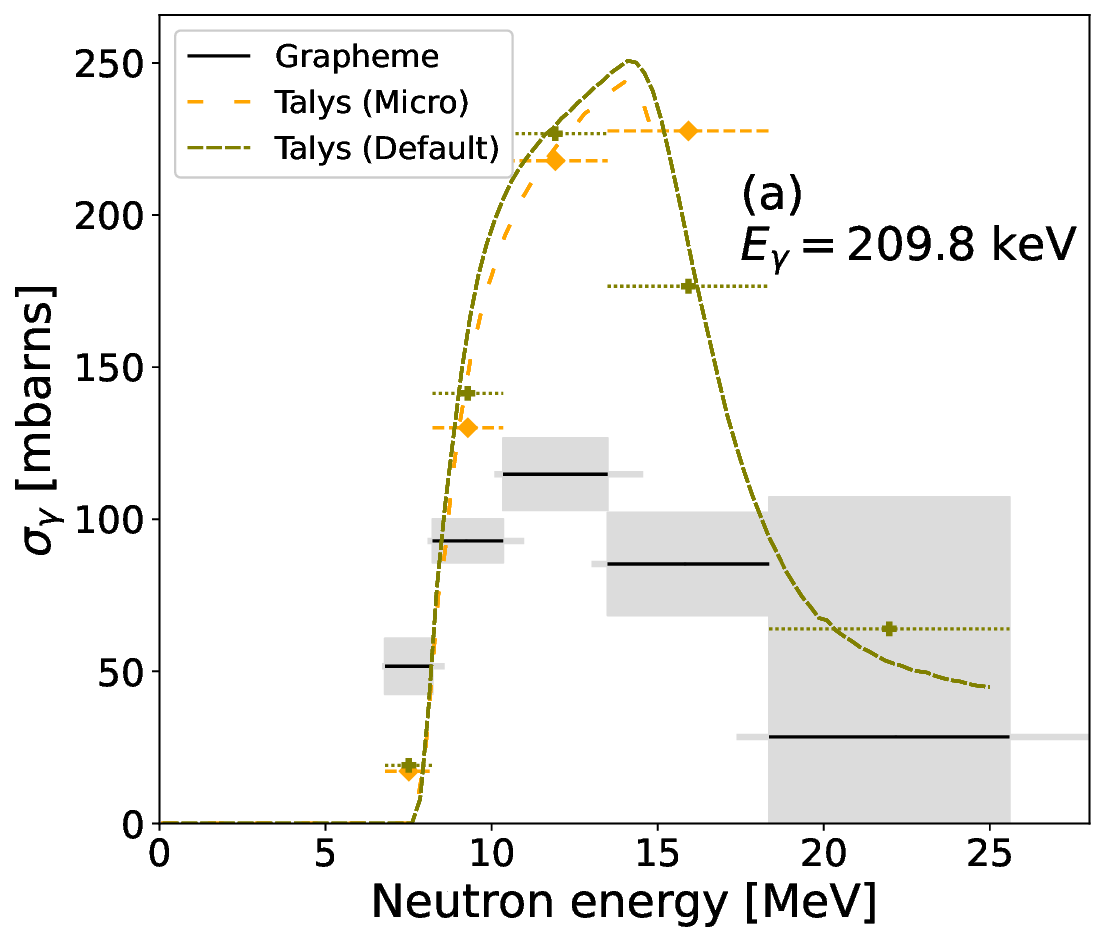}
\includegraphics[scale=.41]{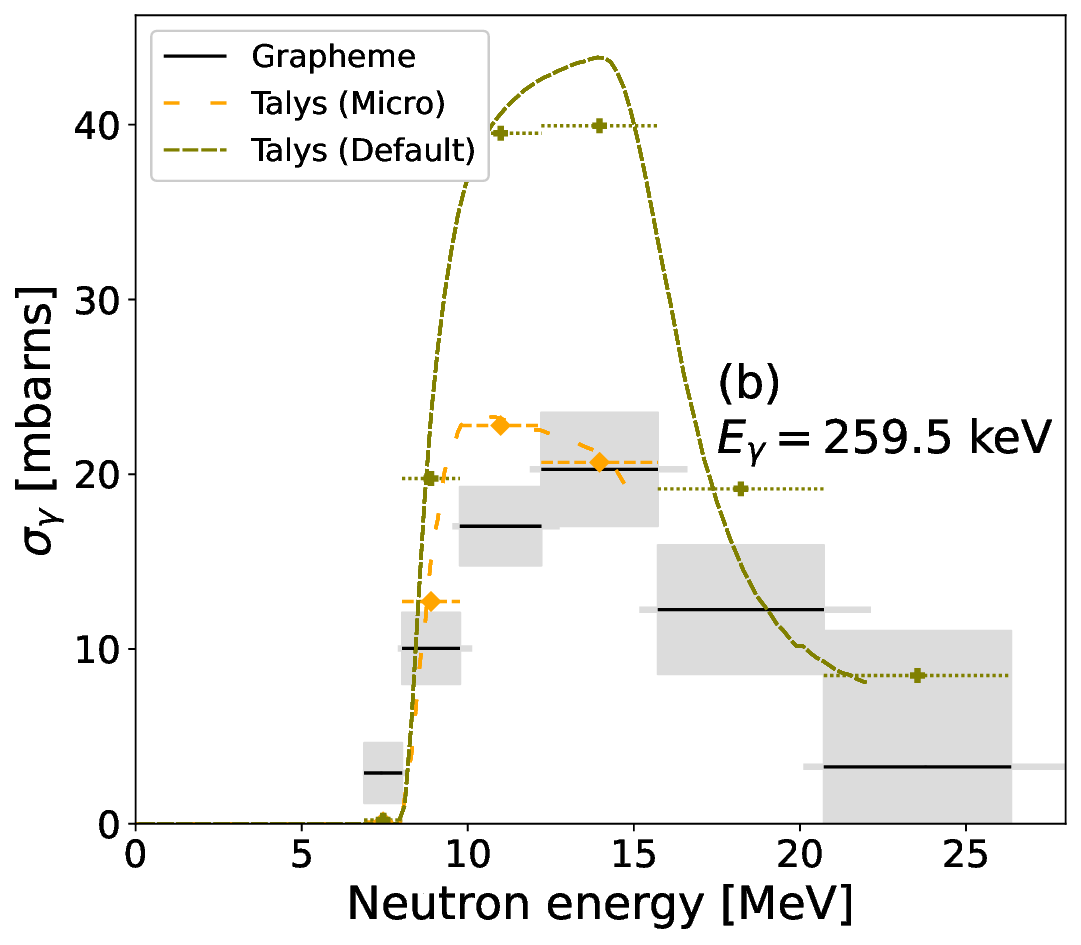}
\includegraphics[scale=.41]{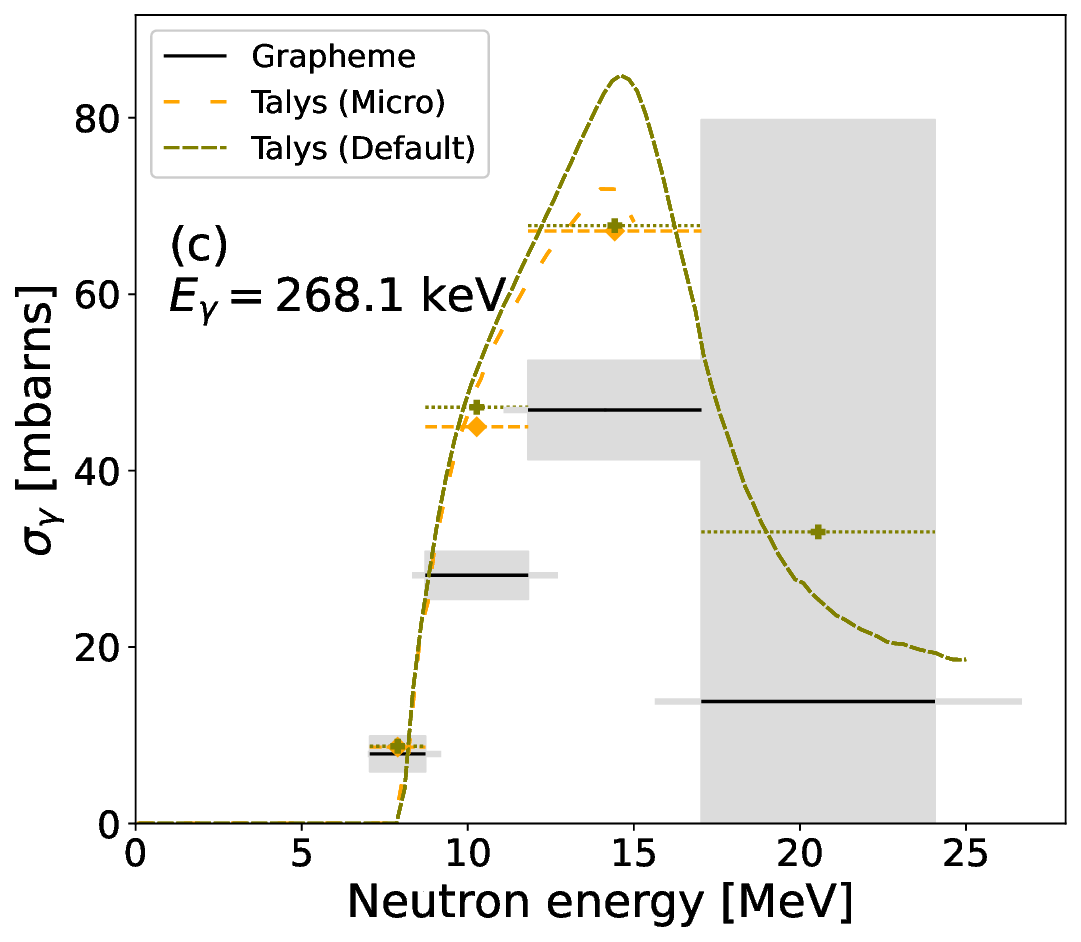}
\includegraphics[scale=.41]{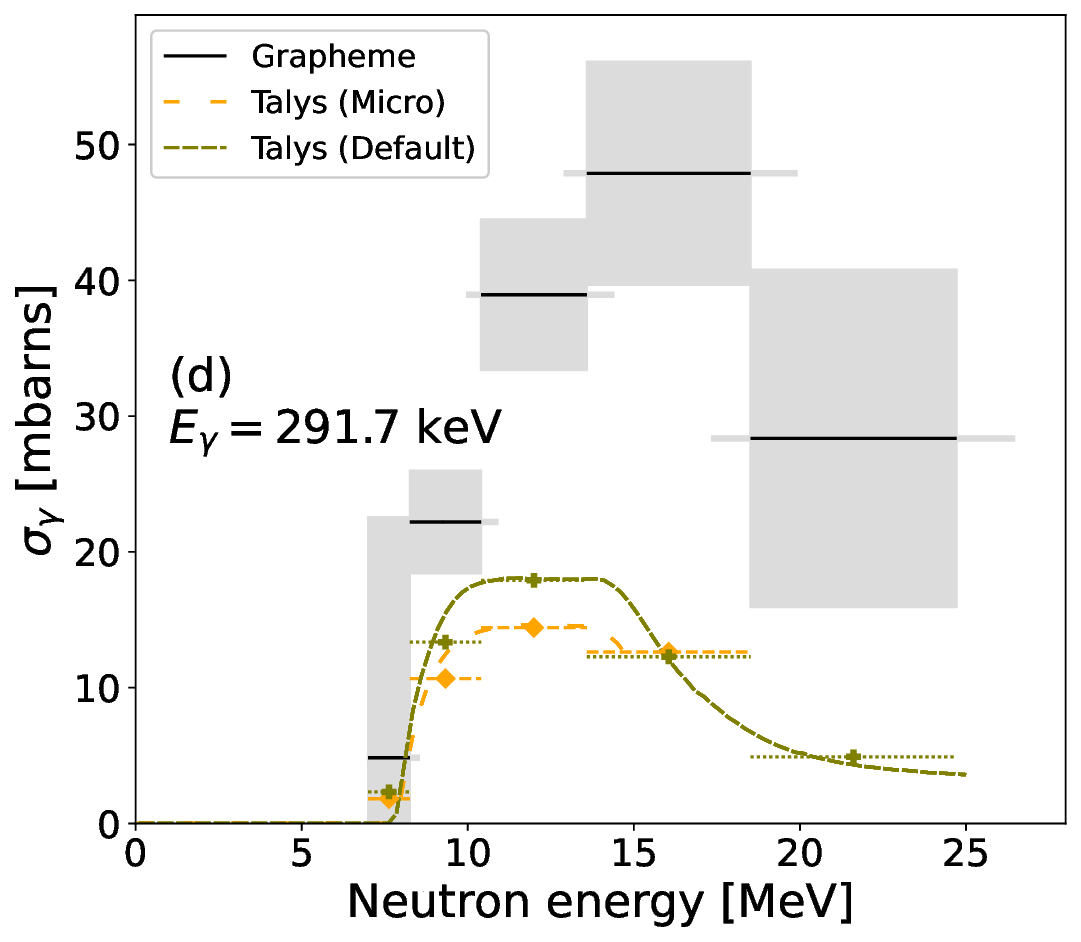}
\includegraphics[scale=.41]{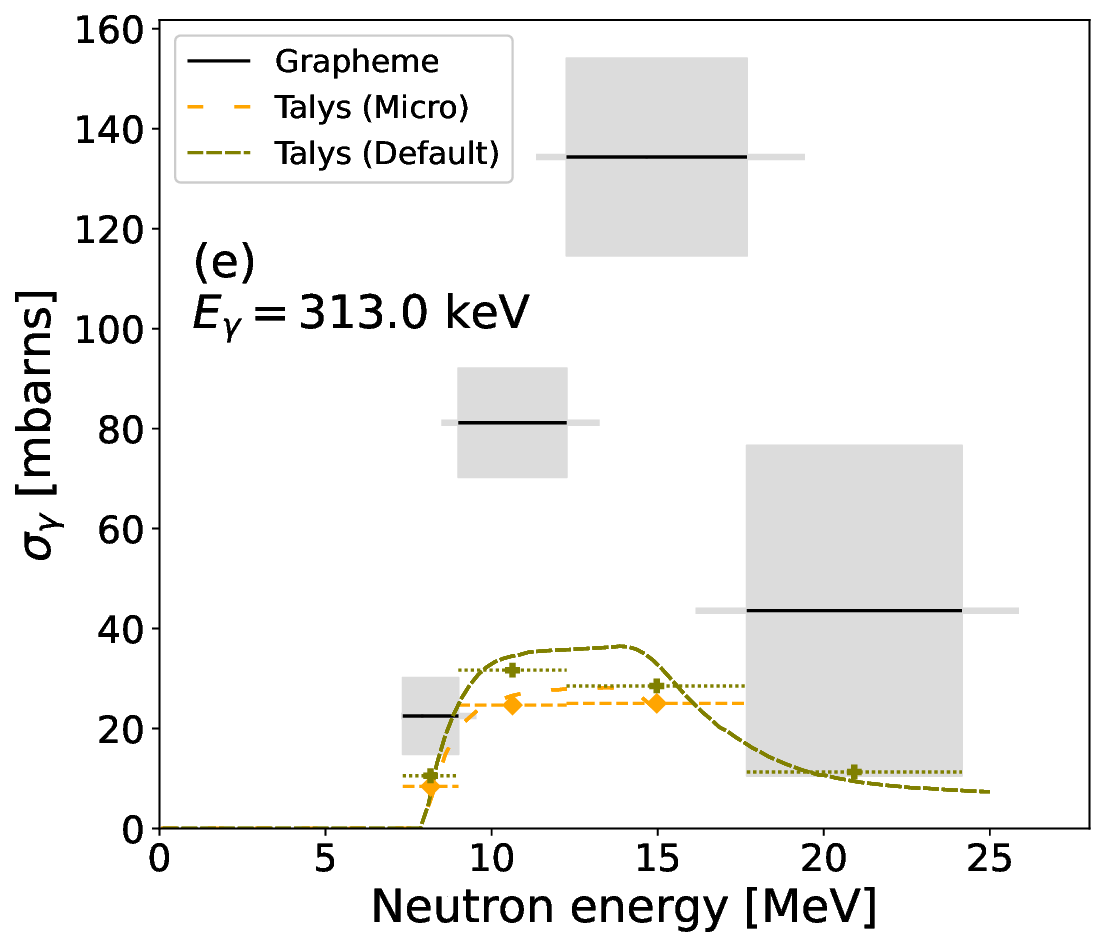}
\includegraphics[scale=.41]{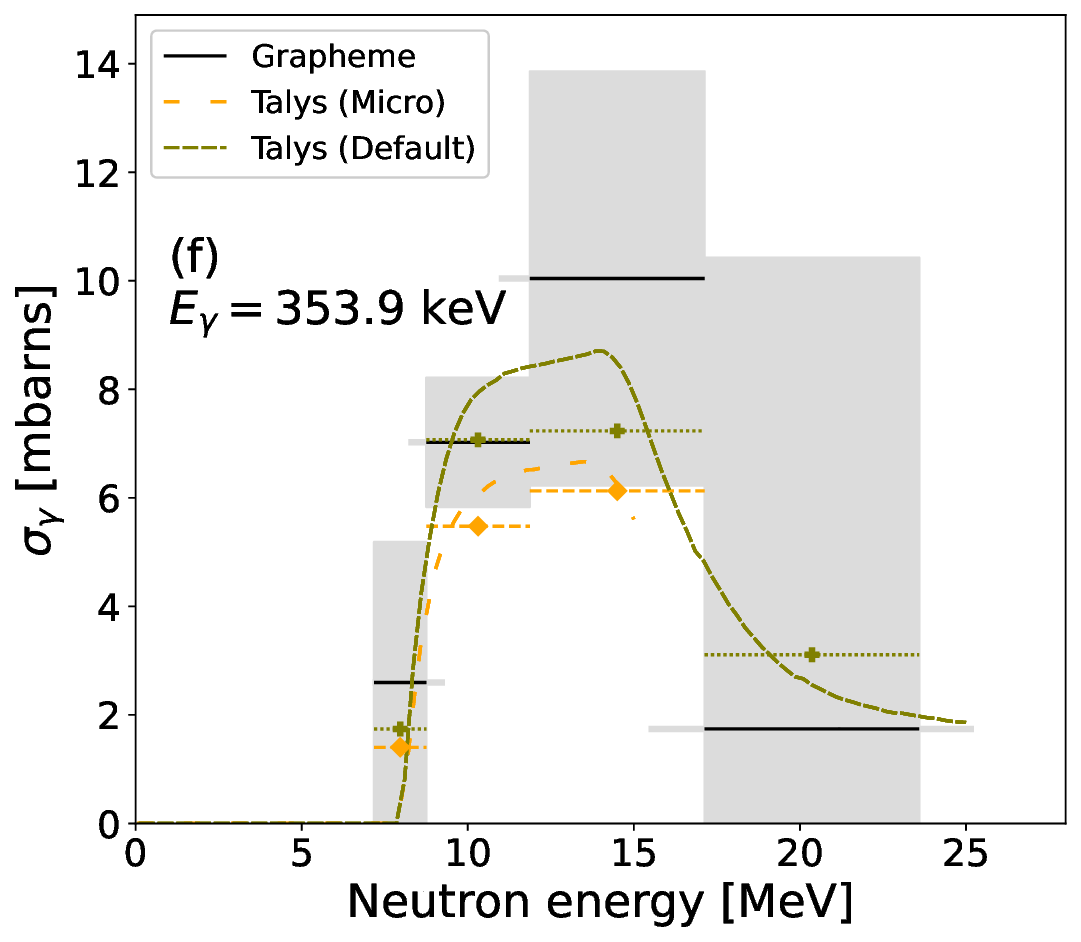}
\caption{
Experimental \Wquatre\ndeuxng{} cross sections for the six studied transitions: (a) 209.8, (b) 259.5, (c) 268.1, (d) 291.7, (e) 313.0 and (f) 353.9~keV.
The horizontal solid black lines indicate the measured cross section in each neutron energy bin, and the surrounding shaded areas the total combined uncertainty.
The small gray lines extending beyond the bins represent the uncertainty on the bin limits ($\delta_{E_\text{n}^\text{low}}$ and $\delta_{E_\text{n}^\text{high}}$ in Tables~\ref{tab:209}--\ref{tab:354}).
The TALYS predictions with default parameters are shown as a short-dashed olive line, with the values averaged over each neutron energy bin shown as $+$ markers and dotted horizontal lines.
The \textit{microscopic} calculations are shown as a long-dashed orange line, with bin-averaged values shown as diamond markers and dashed horizontal lines.
Figure adapted from Ref.~\cite{figshare}.
}
\label{fig:n2nall}
\end{figure*}

Along with tables, we represent the cross sections in graphical form in Fig.~\ref{fig:n2nall}.

For $\gamma$ rays of energies between 200 and 350 keV, the time resolution (which is slightly $E_\gamma$ dependant) is about 35 to 50 ns. 
At \SI{28.8}{m}, that translates into an energy resolution of about 0.4 MeV for neutrons of energies around \SI{7}{MeV}, and about \SI{2}{MeV} for the higher energies ($E_n\approx$ \SI{22}{MeV}). 
This constrains the size of the ranges over which the average experimental cross section is extracted, with the uncertainty on the flight distance and time calibration adding extra uncertainty, visualized in the Fig.~\ref{fig:n2nall} with a small gray line expanding beyond the ranges.

The experimental values are compared, in the figures, to calculations performed with the TALYS nuclear reaction code~\cite{Koning2023}. 
Two computations are plotted on the figures: the first one, using the \texttt{default} parameters (noted "Default" in the figures' labels), and the second using microscopic description in the entrance and preequilibrium stages of the reaction (in the rest of the text referred as \textit{microscopic}, noted "Micro" in the figures, calculations stop at 15~MeV, so they do not extend to high energy range like the "default" ones)~\cite{MDupuisPV}.
In this paper, the calculations are displayed as an example of model results. 
Investigation of how the experimental data can be used to improve the models will be dedicated to a future paper combining data describe in this article, as well as Refs.~\cite{PRCeveneven} and \cite{EPJA183W}.

\begin{table*}[!ht]
    \centering
    \begin{tabular}{rrrrrr}
     \multicolumn{1}{c}{$E_\text{n}^\text{low}$ [keV]} &
     \multicolumn{1}{c}{$\delta_{E_\text{n}^\text{low}}$ [keV]} &
     \multicolumn{1}{c}{$E_\text{n}^\text{high}$ [keV]} &
     \multicolumn{1}{c}{$\delta_{E_\text{n}^\text{high}}$ [keV]} &
     \multicolumn{1}{c}{$\sigma_\gamma$ [mb]} &
     \multicolumn{1}{c}{$u_{\sigma_\gamma}$ [mb]}\\
     \hline
6790.16 & 83.99 & 8222.14 & 364.41 &  51.7 & 9.2 \\ 
8222.14 & 146.26 & 10343.58 & 637.10 &  92.9 & 7.2 \\ 
10343.58 & 263.70 & 13493.57 & 1070.80 &  114.8 & 11.8 \\ 
13493.57 & 482.19 & 18356.31 & 1952.68 &  85.3 & 16.9 \\ 
18356.31 & 980.93 & 25591.26 & 3893.76 &  28.4 & 78.8 \\ 

    \end{tabular}
    \caption{
    Experimental value of the measured \Wquatre\ndeuxng{} cross section for the 209.8~keV transition.
Columns one and two report the lower value of the neutron energy range over which the cross section is extracted , $E_\text{n}^\text{low}$ in keV, and the uncertainty on this value, $\delta_{E_\text{n}^\text{low}}$ (due to time of flight and distance of flight uncertainty).
The next two columns are the same for the upper value of the range (E$_\text{n}^\text{high}$ and $\delta_{E_\text{n}^\text{high}}$.
The fifth column gives the measured $\gamma$ cross section, $\sigma_\gamma$ in millibarns.
The last column gives the total combined (i.e. statistical and systematic) uncertainty on the measured $\gamma$ cross section, $\text{u}_{\sigma_\gamma}$.
    }
    \label{tab:209}
\end{table*}

\begin{table*}[!ht]
    \centering
    \begin{tabular}{rrrrrr}
     \multicolumn{1}{c}{$E_\text{n}^\text{low}$ [keV]} &
     \multicolumn{1}{c}{$\delta_{E_\text{n}^\text{low}}$ [keV]} &
     \multicolumn{1}{c}{$E_\text{n}^\text{high}$ [keV]} &
     \multicolumn{1}{c}{$\delta_{E_\text{n}^\text{high}}$ [keV]} &
     \multicolumn{1}{c}{$\sigma_\gamma$ [mb]} &
     \multicolumn{1}{c}{$u_{\sigma_\gamma}$ [mb]}\\
     \hline
6882.73 & 41.15 & 8021.66 & 207.83 &  2.9 & 1.7 \\ 
8021.66 & 127.47 & 9764.74 & 362.47 &  10.0 & 2.0 \\ 
9764.74 & 227.60 & 12213.06 & 552.71 &  17.0 & 2.2 \\ 
12213.06 & 345.67 & 15722.18 & 890.60 &  20.3 & 3.2 \\ 
15722.18 & 558.72 & 20722.21 & 1415.02 &  12.3 & 3.7 \\ 
20722.21 & 619.23 & 26357.08 & 1631.82 &  3.3 & 7.8 \\ 

    \end{tabular}
        \caption{Same as Table \ref{tab:209}, for the 259.5~keV transition.}

    \label{tab:259}
\end{table*}

\begin{table*}[!ht]
    \centering
    \begin{tabular}{rrrrrr}
     \multicolumn{1}{c}{$E_\text{n}^\text{low}$ [keV]} &
     \multicolumn{1}{c}{$\delta_{E_\text{n}^\text{low}}$ [keV]} &
     \multicolumn{1}{c}{$E_\text{n}^\text{high}$ [keV]} &
     \multicolumn{1}{c}{$\delta_{E_\text{n}^\text{high}}$ [keV]} &
     \multicolumn{1}{c}{$\sigma_\gamma$ [mb]} &
     \multicolumn{1}{c}{$u_{\sigma_\gamma}$ [mb]}\\
     \hline
7054.32 & 65.26 & 8723.83 & 473.18 &  7.9 & 2.0 \\ 
8723.83 & 401.23 & 11817.24 & 894.48 &  28.1 & 2.7 \\ 
11817.24 & 739.87 & 17035.16 & 1697.58 &  46.9 & 5.6 \\ 
17035.16 & 1406.01 & 24064.27 & 2616.97 &  13.8 & 65.9 \\ 

    \end{tabular}
        \caption{Same as Table \ref{tab:209}, for the 268.1~keV transition.}

    \label{tab:268}
\end{table*}

\begin{table*}[!ht]
    \centering
    \begin{tabular}{rrrrrr}
     \multicolumn{1}{c}{$E_\text{n}^\text{low}$ [keV]} &
     \multicolumn{1}{c}{$\delta_{E_\text{n}^\text{low}}$ [keV]} &
     \multicolumn{1}{c}{$E_\text{n}^\text{high}$ [keV]} &
     \multicolumn{1}{c}{$\delta_{E_\text{n}^\text{high}}$ [keV]} &
     \multicolumn{1}{c}{$\sigma_\gamma$ [mb]} &
     \multicolumn{1}{c}{$u_{\sigma_\gamma}$ [mb]}\\
     \hline
7001.01 & 37.01 & 8254.38 & 306.65 &  4.8 & 17.7 \\ 
8254.38 & 269.90 & 10398.90 & 522.18 &  22.2 & 3.8 \\ 
10398.90 & 447.35 & 13583.26 & 827.76 &  38.9 & 5.5 \\ 
13583.26 & 699.71 & 18512.15 & 1418.21 &  47.9 & 8.2 \\ 
18512.15 & 1186.57 & 24714.23 & 1762.77 &  28.4 & 12.4 \\ 

    \end{tabular}
        \caption{Same as Table \ref{tab:209}, for the 291.7~keV transition.}

    \label{tab:291}
\end{table*}

\begin{table*}[!ht]
    \centering
    \begin{tabular}{rrrrrr}
     \multicolumn{1}{c}{$E_\text{n}^\text{low}$ [keV]} &
     \multicolumn{1}{c}{$\delta_{E_\text{n}^\text{low}}$ [keV]} &
     \multicolumn{1}{c}{$E_\text{n}^\text{high}$ [keV]} &
     \multicolumn{1}{c}{$\delta_{E_\text{n}^\text{high}}$ [keV]} &
     \multicolumn{1}{c}{$\sigma_\gamma$ [mb]} &
     \multicolumn{1}{c}{$u_{\sigma_\gamma}$ [mb]}\\
     \hline
7325.19 & 50.26 & 8996.96 & 547.15 &  22.5 & 7.6 \\ 
8996.96 & 512.50 & 12256.93 & 996.97 &  81.2 & 10.9 \\ 
12256.93 & 914.39 & 17686.72 & 1747.41 &  134.3 & 19.7 \\ 
17686.72 & 1548.51 & 24154.65 & 1721.22 &  43.6 & 33.0 \\ 

    \end{tabular}
        \caption{Same as Table \ref{tab:209}, for the 313.0~keV transition.}

    \label{tab:313}
\end{table*}

\begin{table*}[!ht]
    \centering
    \begin{tabular}{rrrrrr}
     \multicolumn{1}{c}{$E_\text{n}^\text{low}$ [keV]} &
     \multicolumn{1}{c}{$\delta_{E_\text{n}^\text{low}}$ [keV]} &
     \multicolumn{1}{c}{$E_\text{n}^\text{high}$ [keV]} &
     \multicolumn{1}{c}{$\delta_{E_\text{n}^\text{high}}$ [keV]} &
     \multicolumn{1}{c}{$\sigma_\gamma$ [mb]} &
     \multicolumn{1}{c}{$u_{\sigma_\gamma}$ [mb]}\\
     \hline
7177.69 & 40.11 & 8750.10 & 556.84 &  2.6 & 2.6 \\ 
8750.10 & 535.47 & 11867.48 & 986.59 &  7.0 & 1.2 \\ 
11867.48 & 933.42 & 17132.80 & 1791.61 &  10.0 & 3.8 \\ 
17132.80 & 1686.92 & 23592.63 & 1643.96 &  1.7 & 8.7 \\ 

    \end{tabular}
        \caption{Same as Table \ref{tab:209}, for the 353.9~keV transition.}

    \label{tab:354}
\end{table*}

\section{Conclusion}

New experimental cross sections for six discrete $\gamma$-ray transitions following the \Wquatre\ndeuxn{} reaction have been measured.
Limitations include the neutron energy resolution (\SIrange{0.4}{2}{MeV}) and the statistical uncertainty at high energy due to the low yields.
Nevertheless, these results fill a significant gap in the existing experimental nuclear data for tungsten isotopes, providing crucial information for both fundamental nuclear physics and applied nuclear technology.
This data set, alongside those obtained for other tungsten isotopes~\cite{PRCeveneven,EPJA183W}, provides valuable new benchmarks for the validation and improvement of nuclear reaction models, contributing to enhanced accuracy of evaluated nuclear data libraries.

\section*{Data Availability}

The data associated with this article are available in the EXFOR database Entry \texttt{23104.008} \url{https://www-nds.iaea.org/EXFOR/23104} \cite{result184Wexfor}, on the \href{https://entrepot.recherche.data.gouv.fr/}{Recherche Data Gouv} repository platform \cite{QFFRD8_2025}, and as a git repository \cite{result184Wgit}, under the CC-BY-4.0 Open Access Licence.

    "Experimental (n,~2n~$\gamma$) cross sections for $^{184}$W"
    HENNING, Greg, \textit{et al.}, 2025, Recherche Data Gouv,
    \url{https://doi.org/10.57745/QFFRD8}.

\section*{Acknowledgements} 

The authors thank the team at the GELINA facility and IPHC for their
support.
This work was funded in part by NEEDS, and was conducted within the EEDIN collaboration\cite{EEDINHAL}. 
It  was also supported, in parts, by the PACEN and GEDEPEON programs, and by the European Commission within the Sixth Framework Program through I3-EFNUDAT (EURATOM contract No. 036434) and NUDAME (Contract FP6-516487), within the Seventh Framework Program through EUFRAT (EURATOM contract No. FP7-211499), through ANDES (EURATOM contract No. FP7-249671), and from the Euratom research and training program 2014-2018 under grant agreement No. 847552 (SANDA).
K.H.~Guber was supported by the DOE Nuclear Criticality Safety Program, funded and managed by the National Nuclear Security Administration for the US Department of Energy.
T.~Kawano carried out this work under the auspices of the National Nuclear Security Administration of the U.S. Department of Energy at Los Alamos National Laboratory under Contract No. 89233218CNA000001.

\bibliographystyle{elsarticle-num-names}       
\bibliography{paper}   

\end{document}